# Measurement of mobility in dual-gated MoS$_2$ transistors

**To the Editor** – Atomically thin semiconducting MoS$_2$ is of great interest for high-performance flexible electronic and optoelectronic devices. Initial measurements using back-gated field-effect transistor (FET) structures on SiO$_2$ yielded mobility of 1-50 cm$^2$/Vs for few-layer MoS$_2$[1, 2]. However, greatly increased mobility – as high as 900 cm$^2$/Vs – was recently reported for monolayer MoS$_2$ by Radisavljevic, *et al.*[Ref. 3; see also Refs. 4-6], and for multilayer MoS$_2$ by others[7, 8], in devices covered by a high-κ dielectric layer and a metal top gate. Similar increases were reported for electrolyte top-gated MoS$_2$[9]. A puzzling common aspect of these reports is that the double-gated devices consistently show much higher mobility when measured using the weakly-coupled back gate (typically 300 nm of SiO$_2$) than using the strongly-coupled top gate. While this difference might be attributed to a different quality of the top and bottom surfaces, the large discrepancy motivates close examination of the measurement techniques to rule out spurious effects. Notably, it has been shown that the mobility may be greatly overestimated in this geometry[10, 11]. There is good evidence that this is the case here[4-9]; below we discuss Radisavljevic, *et al.*[4] as an illustrative example.

In an FET in the linear regime, the charge carrier density is controlled by capacitive coupling to a gate, so that the drain current $I_{ds}$ is given by

$$I_{ds} = \frac{\mu C}{L^2}(V_g - V_{th})V_{ds} \qquad (1)$$

where μ is the field-effect mobility, $V_g$ the gate voltage, $V_{th}$ the threshold voltage, $V_{ds}$ the drain voltage, $L$ the channel length, and $C$ the gate-channel capacitance. This leads to an approximate expression for μ:

$$\mu = \frac{L^2}{CV_{ds}}\frac{dI_{ds}}{dV_g}. \qquad (2)$$

Correct measurement of mobility using Eqn. (2) requires knowledge of the gate capacitance. In the double-gated geometry, capacitive coupling between the back and top gates through the large top-gate bonding pad can cause the back-gate capacitance to deviate significantly from the nominal parallel-plate value[11]. Figure 1a shows a schematic of the double-gated device, with top-gate capacitance $C_{tg}$, back gate capacitance $C_{bg}$, and coupling $C_{bt}$ between back and top gates. In a device with thin high-κ top gate dielectric and ~ (100 μm)$^2$ bonding pads, $C_{bt} \gg C_{tg} \gg C_{bg}$. If the device is measured by sweeping the back gate with the top gate electrode disconnected, then the effective back-gate capacitance $C = C_{bg} + (C_{tg}^{-1} + C_{bt}^{-1})^{-1} \approx C_{tg}$. Neglecting this coupling would lead to an over-estimation of the mobility by a factor $C_{tg}/C_{bg}$. This experimental artifact has been observed before in graphene top-gated transistors[10, 11].

To illustrate the difficulty in estimating the mobility using a double-gated structure, in Figure 1b we show three data sets extracted from Radisavljevic, *et al.*[4], measured on the same device. The blue curve shows the current vs. back gate voltage with top gate disconnected; using this data Radisavljevic, *et al.*[4] assumed $C = C_{bg}$ and extracted a mobility of 217 cm$^2$/Vs, however $C$ is unknown in this setup because it includes spurious coupling through the top gate. The green data shows the current vs. top gate voltage. Remarkably, even though the top gate capacitance per unit area is ~ 43× larger, the two curves are nearly identical in $V_{th}$ and slope, suggesting strong coupling of back and top gate. Here Eq. (2), with $L$ the top gate length and $C = C_{tg}$ (a good approximation since $C_{tg} \gg C_{bg}$), yields μ ≈ 2 cm$^2$/Vs. The current as a function of back gate voltage, with $V_{tg}$ fixed at 0, is shown in red, from which we calculate μ ≈ 7 cm$^2$/Vs; in this geometry $C = C_{bg}$ is well-defined. Hence top- and back-gated measurements in setups with well-defined capacitances yield μ = 2-7 cm$^2$/Vs, comparable to previously reported back-gate-only measurements[1, 2] (the discrepancy between the two values may reflect different series resistances as the top gate does not gate the entire device).

While it is possible that Eqn. 2 underestimates the true channel mobility due to e.g. contact resistance, we find no evidence in Refs. 3-9 that the mobility is significantly higher than ~10 $cm^2$/Vs. The true electronic performance of $MoS_2$, as well as the possible role of applied dielectrics in improving mobility, remains to be accurately determined. Rigorous measurement will require separation of contact and channel effects, e.g. by four-terminal or transmission line methods, and careful determination of the capacitive coupling, e.g. by Hall effect.

**Michael S. Fuhrer[1,2] and James Hone[3]**
[1]Center for Nanophysics and Advanced Materials, Department of Physics, University of Maryland, College Park, MD 20742 USA;
[2]School of Physics, Monash University, Melbourne VIC 3800, Australia;
[3]Department of Mechanical Engineering, Columbia University, New York, New York 10027, USA.
*email:mfuhrer@umd.edu; jh2228@columbia.edu.

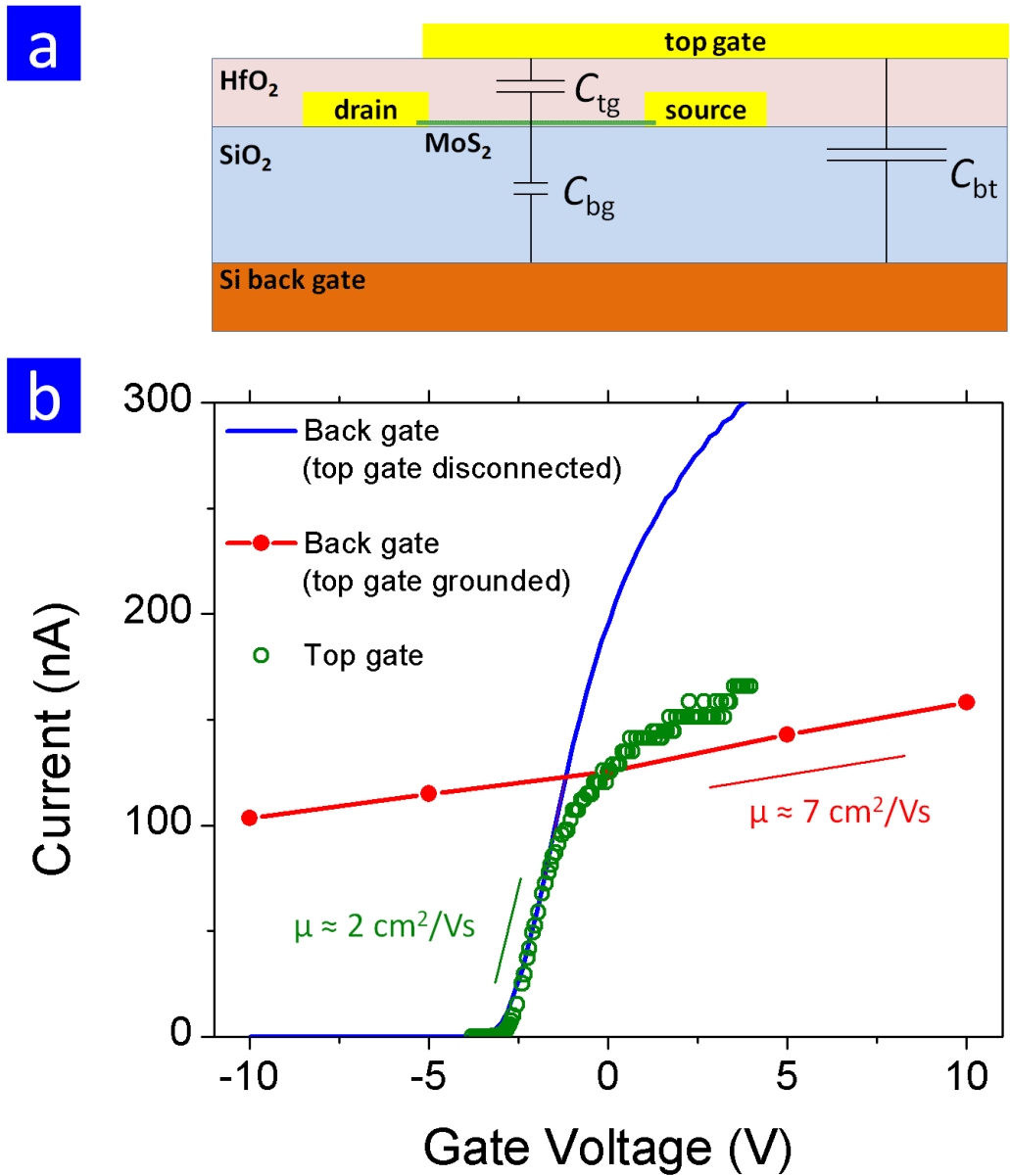

Figure 1. (a) Schematic of MoS$_2$ dual-gated transistor in Ref. [3] showing capacitances discussed in text. Drawing is not to scale. (b) Drain current vs. gate voltage for single-layer MoS$_2$ transistor with gate voltage applied to back gate with top gate disconnected (solid blue line), back gate with top gate grounded (red circles), or top gate (green circles). Drain voltage is 10 mV in all cases. Data are taken from Ref. [3], Figs. 3b and Fig. 4a.